\documentclass[pre,twocolumn]{revtex4}
\usepackage{url}
\usepackage{color}
\usepackage{graphicx}
\usepackage{amsmath}
\usepackage{algorithm,algpseudocode}

\begin{document}

\title{SIMD Vectorization for the Lennard-Jones Potential with AVX2 and AVX-512 instructions}

\author{Hiroshi Watanabe}
\thanks{Corresponding author}
\email{hwatanabe@issp.u-tokyo.ac.jp}
\author{Koh M. Nakagawa}
\affiliation{
The Institute for Solid State Physics, The University of Tokyo,
Kashiwanoha 5-1-5, Kashiwa, Chiba 277-8581, Japan
}

\date{\today}


\begin{abstract}
This work describes the SIMD vectorization of the force calculation of the Lennard-Jones potential with Intel AVX2 and AVX-512 instruction sets. Since the force-calculation kernel of the molecular dynamics method involves indirect access to memory, the data layout is one of the most important factors in vectorization. We find that the Array of Structures (AoS) with padding exhibits better performance than Structure of Arrays (SoA) with appropriate vectorization and optimizations. In particular, AoS with 512-bit width exhibits the best performance among the architectures. 
While the difference in performance between AoS and SoA is significant for the vectorization with AVX2, that with AVX-512 is minor. The effect of other optimization techniques, such as software pipelining together with vectorization, is also discussed. We present results for benchmarks on three CPU architectures: Intel Haswell (HSW), Knights Landing (KNL), and Skylake (SKL). The performance gains by vectorization are about 42\% on HSW compared with the code optimized without vectorization. On KNL, the hand-vectorized codes exhibit 34\% better performance than the codes vectorized automatically by the Intel compiler. On SKL, the code vectorized with AVX2 exhibits slightly better performance than that with vectorized AVX-512.
\end{abstract}

\maketitle

\section{Introduction}

Since Alder and Wainwright performed molecular dynamics (MD) simulations for the first time~\cite{Alder1957},
MD has been an important tool for exploring the wide variety of fields in science.
Starting from the first MD simulation with 32 atoms, the number of atoms involved in MD has continued to increase owing to the development of computational power and has reached hundreds of billions to trillions of atoms~\cite{GERMANN2008,Watanabe2013}. It is expected that the size and complexity of  systems will continue to increase with the development of computers. Since the increase in the CPU frequency stopped in the early 2000s~\cite{Rupp2018}, the performance development of CPUs has mainly depended on increasing the number of CPU cores and the SIMD width. SIMD, which is short for single instruction multiple data, enables us to process multiple data with a single instruction. The width of SIMD is the number of data that can be processed simultaneously and it is determined by the bit length of registers. Beginning with SSE, which supports 128-bit registers, 256-bit registers for AVX and 512-bit registers for AVX-512 are available, where SSE and AVX are shorts for Streaming SIMD extensions and advanced vector extensions, respectively. Since the theoretical peak performance of a CPU is the value when the application uses the vector width to the full, the utilization of SIMD is crucial for the performance of applications on modern CPU architecture. 
However, it is not trivial to transform a scalar kernel to a SIMD-vectorized one. Additionally, the optimal method of vectorization is different for each instruction set architecture (ISA). 
Portability is one of the important issues for vectorization.
The SIMD capabilities are usually provided as an extension of the existing ISA. SIMD and related instructions are different for each architecture.
To address this issue, several approaches have been proposed.
One of them is a directive-based approach~\cite{Brown2015}.
By using directives, one can utilize GPGPU or many-core processors effectively while
retaining portability.
Another is a framework approach. Karpi\'{n}ski and McDonald provided
a framework called Unified Multi/Many-Core Environment (UME), which allows
a programmer to utilize the SIMD capabilities without detailed knowledge of the specific architecture~\cite{Karpinski2017}. With these approaches, one can develop an application
that exhibits high performance without increasing software complexity.

Despite the above efforts, SIMD vectorization remains a difficult and cumbersome task, since the performance would be unsatisfactory simply by vectorizing the existing scalar code.
To utilize the SIMD capability of modern CPUs, it is necessary to combine SIMD vectorization with an optimal data layout and other optimization techniques.
In this paper, we describe the SIMD vectorization of the force calculation for the Lennard-Jones (LJ) potential with AVX2 and AVX-512 on several types of CPU. The force calculation is the most time-consuming part of MD, and therefore, the efficiency of the kernel directly determines the performance of MD.
Since the force-calculation kernel is a typical example of the vectorization of a loop with indirect access, there have been many reports on the vectorization of MD codes, including that of the many-body potential~\cite{Hohnerbach2016}.
Additionally, the important kernels of widely used MD packages, such as Gromacs~\cite{Abraham2015}, LAMMPS~\cite{Edwards2014}, and NAMD~\cite{Phillips2005}, have already been effectively vectorized and are available.
However, the previous works mainly focused on the performance of the whole application instead of that of the vectorized kernel. While the performance of the application is ultimately most important, the total performance depends on various factors such as the communication and the cost of thread synchronization.
Such factors make it difficult to evaluate the impact of vectorization. Therefore, we focus on force-calculation kernel in this paper.In particular, we investigate in detail the impact of optimization techniques when used with vectorization. We also discuss the difference between the vectorization with AVX2 and AVX-512 and between CPU architectures.
Since the computational intensity of LJ is low, \textit{i.e.}, a number of floating point operations is small relative to the amount of memory access, the memory access is the main bottleneck.
Therefore, most of our efforts are devoted to optimizing memory access.
The optimization techniques described in this paper can be applied to other cases with low computational intensity. Although mixed-precision calculations are known to improve the performance of MD, we use double-precision numbers for all calculations throughout this work for simplicity.

The rest of the article is organized as follows. 
The benchmark conditions are described in the next section.
Then we give a brief introduction to the optimization techniques employed before vectorization in Sec.~\ref{sec:scalar}.
Vectorization with AVX2 is described in Sec.~\ref{sec:avx2}
and that with AVX-512 and related optimizations are described in Sec.~\ref{sec:avx512}.
The benchmark results on SKL are shown in Sec.~\ref{sec:skl}.
Section~\ref{sec:summary} is devoted to a summary and discussion.
The associated code is available at~\cite{lj_simd}.

\section{Benchmark Conditions} \label{sec:benchmark}

We consider three types of CPU architecture; Intel Haswell (HSW), Knights Landing (KNL), and Skylake (SKL), which we refer to as HSW, KNL, and SKL, respectively.
We performed benchmark simulations on the systems of the Institute for Solid State Physics of the University of Tokyo for HSW and SKL and of the Information Technology Center of the University of Tokyo for KNL.
The conditions for the benchmark simulations are as follows.
The simulation box is a cube with a linear size of 100 $\sigma$, where $\sigma$ is the diameter of an LJ atom.
Atoms are placed at the face-centered-cubic lattice.
The cutoff length is 3.0 $\sigma$.
The number of atoms is 119164, and consequently, the number density is 1.0.
Only force calculations are performed and the time to calculate the force 100 times is measured.
The program is executed as a single-threaded process on a single CPU core.
The software is compiled using Intel C++ compiler.
The details of the CPU, the versions of the compiler, and the compilation options are listed in Table~\ref{tbl:CPUs}.

\begin{table*}
\begin{tabular}{llllll}
Name & Processor &  Vector ISA  & Compiler & Options \\
\hline
HSW & Intel Xeon E5-2680 v3 & AVX2 & icpc (ICC) 16.0.4& -xHOST \\
KNL & Intel Xeon Phi 7250  & AVX2, AVX-512 & icpc (ICC) 18.0.1 & -axMIC-AVX512\\
SKL & Intel Xeon Gold 6148 & AVX2, AVX-512 & icpc (ICC) 18.0.1& -xCORE-AVX512 -qopt-zmm-usage=high
 \\
\end{tabular}
\caption{%
Details of CPUs, versions of the compiler, and compilation options used for benchmark simulations.
The common compiler option is {\ttfamily -O3 -std=c++11 -w2 -w3 -diag-disable:remark -restrict} and
only CPU-dependent options are shown.
}\label{tbl:CPUs}
\end{table*}

\begin{table}
\begin{tabular}{lp{8cm}}
Name & Description \\
\hline
Pair & Naive implementation using a pair list. \\
Sorted & Reduce the memory access by sorting a list. \\
AoS-4 & Adopt the AoS data structure with padding. 
\end{tabular}
\caption{List of optimizations.
}
\end{table}

\section{Optimization before vectorization} \label{sec:scalar}

\begin{figure}[ht]
\includegraphics[width=7cm]{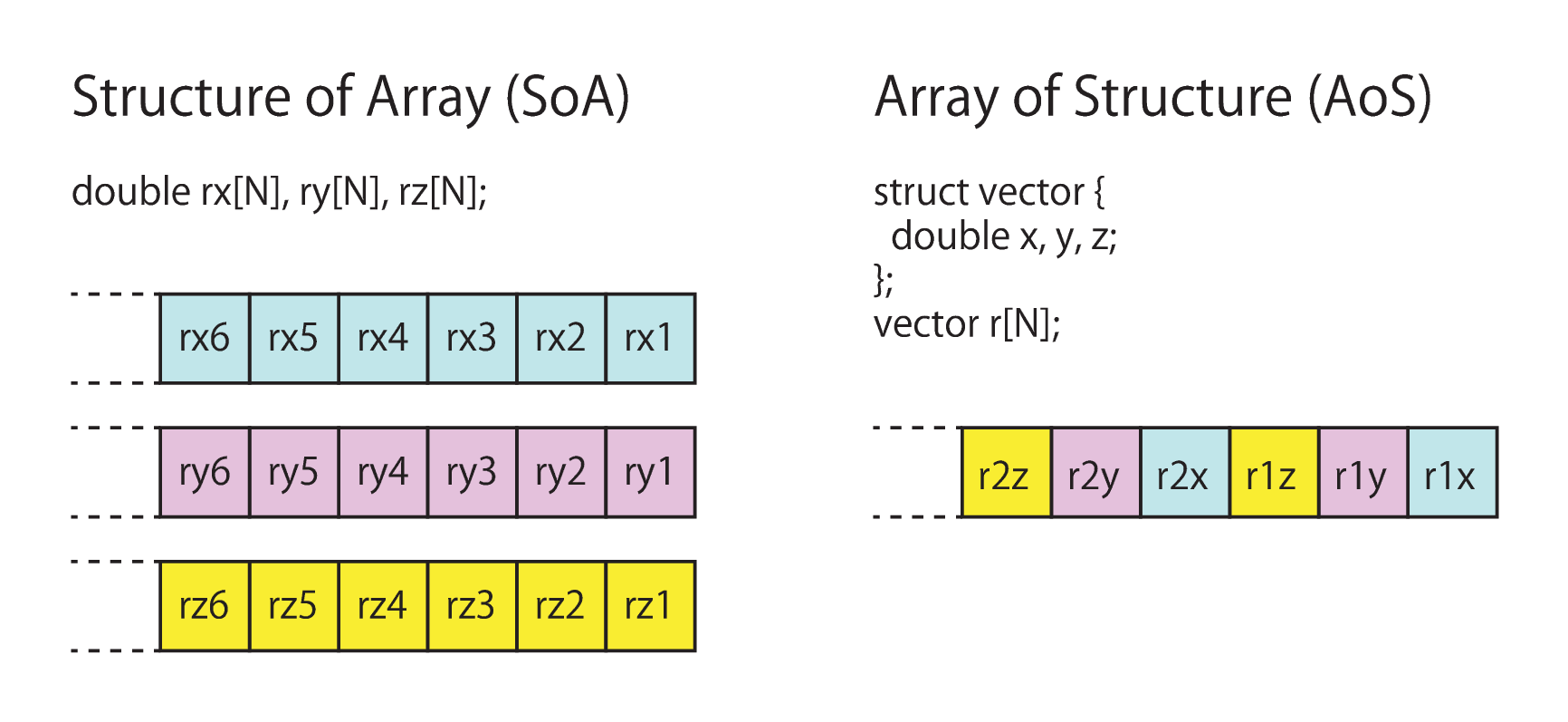}
\caption{
(Color online) Data layout. The numbers represent the indices of atoms.
Left: Structure of Array (SoA). The $x$ coordinates of the atoms are contiguously arranged in the memory. The same is true for the $y$ and $z$ coordinates.
Right: Array of Structure (AoS). Coordinates are stored for each atom.
Examples of data definition in C language are also shown for both cases.
}
\label{fig:aossoa}
\end{figure}

\begin{figure}[ht]
\includegraphics[width=7cm]{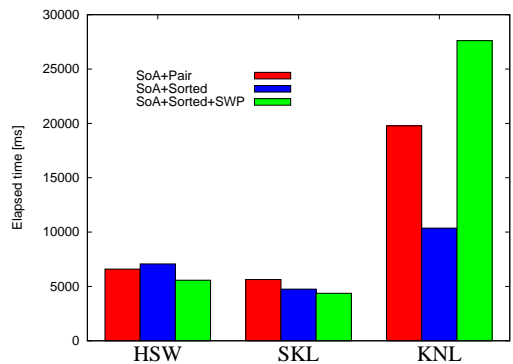}
\caption{
(Color online) Effects of optimizations before vectorization.
}
\label{fig:scalar}
\end{figure}

In this section, we describe several optimization techniques employed
before SIMD vectorization since these techniques affect the SIMD vectorization.
While some of the techniques have been described in a past paper~\cite{Watanabe2011}, we discuss them here to make this paper self-contained. 
Since we consider a three dimensional system, a position vector and velocity vector each contain three elements.
When a system contains $N$ atoms, we have to store $N$ position and velocity vectors in memory.
There are two ways to arrange such data in memory, Structure of Array (SoA) and Array of Structure (AoS).
See Fig.~\ref{fig:aossoa} for the difference between the two layouts.
Here, we adopt SoA as the data layout, \textit{i.e.},
the data are arranged so that the $x$-coordinates of the $i$-atom and $(i+1)$-atom are adjacent.
The results of optimizations are shown in Fig.~\ref{fig:scalar}.
We describe each optimization below.

\subsection{Lennard-Jones Potential}

We consider a classical MD simulation of the LJ potential with truncation. The force calculation of the two-body potential consists of double loops. The loop counter of the outer loop is denoted by $i$ and that of the inner loop is denoted by $j$. The loop counters $i$ and $j$ correspond to the indices of atoms.
Consider the atom pair $(i,j$), for which we calculate the force between them. We call the atom with index $i$ the $i$-atom and the other the $j$-atom.
The coordinates of the $i$- and $j$-atoms are denoted by $\vec{q}_i$ and $\vec{q}_j$, respectively,
and the distance between them is given by $r = |\vec{q}_j - \vec{q}_i|$.
Then the LJ potential is described by
\begin{equation}
V(r) = 4 \varepsilon \left[ \left(\frac{\sigma}{r}\right)^{12}-\left(\frac{\sigma}{r}\right)^6 \right], \label{eq:lj}
\end{equation}
where $\varepsilon$ is the well depth and $\sigma$ is the atomic diameter.
Hereafter, we set $\sigma$ and $\varepsilon$ to unity.
Since the potential decays rapidly as the distance increases,
it is wasteful to calculate the force between pairs at long distances.
Therefore, a cutoff distance is introduced and 
only the interactions of pairs within the distance are considered.
In this paper, we adopt the simple truncation, \textit{i.e.}, 
we introduce a cutoff distance $r_c$ and we ignore force between pairs of atoms at a distance greater than $r_c$.
This truncation modifies the potential function by adding a constant so that $V(r_c) = 0$.
Note that this simple method may exhibit problems in terms of energy conservation.
For practice use, it is better to adopt a truncation method
with not only the potential but also the force becoming zero at the truncation distance~\cite{Stoddard1973, Broughton1983,Holian1991}.
While we here adopt the simple truncation, the optimization and vectorization techniques described in this paper can be applied to other truncation methods with minor modifications.

\begin{algorithm}[tb]
\caption{Force calculation of LJ potential}
\label{alg:lj_force}
\begin{algorithmic}[1]
\State $ \vec{r} \leftarrow \vec{q}_j - \vec{q}_i$
\State $ r^2 \leftarrow \vec{r} \cdot \vec{r}$
\State $ r^6 \leftarrow r^2 \times r^2 \times r^2$
\State $ r^{14} \leftarrow r^6 \times r^6 \times r^2$
\State $ df \leftarrow (48-24\times r^6) \times dt/ r^{14}$
\State $ \vec{p}_i \leftarrow \vec{p}_i - df \times \vec{r} $
\State $ \vec{p}_j \leftarrow \vec{p}_j - df \times \vec{r} $
\end{algorithmic}
\end{algorithm}

The force calculation for a single pair is shown in Algorithm~\ref{alg:lj_force},
for which we count the number of arithmetic operations.
For example, the operation $\vec{q}_j - \vec{q}_i$ involves three subtractions.
The total number of arithmetic operations required to calculate the force between a single pair is 27 addition/multiplications and one division. To calculate the force between a single pair, 
48 bytes must be read and 24 bytes must be written since it is necessary to load the coordinates and momenta of the atoms and write back the momenta.
Considering the above, the force calculation of the LJ system becomes memory-bound.

\subsection{Verlet List and Bookkeeping Method}

Since the potential function is truncated, we must
construct a pair list including only pairs for which 
the distance is shorter than the cutoff distance.
While the trivial implementation to construct the list has a complexity of $O(N^2$) for the total number of atoms $N$,
the complexity of the computation can be reduced to $O(N)$ by adopting the linked-list method~\cite{Quentrec1973,Hockney1988}.
In the construction of the pair list, we register each pair within a search length $r_s$, which is set at longer than $r_c$. Then the list has a margin of $r_s - r_c$ and we can reuse the list for several time steps.
This technique is called the bookkeeping method, which greatly reduces the time required to construct the list~\cite{Verlet1967}.
By monitoring the fastest atom in the system, we can safely determine whether we can continue to use the list or whether we must rebuild it~\cite{Isobe1999}. By adopting the bookkeeping method, pairs beyond the cutoff distance can be registered in the pair list. Therefore, we must exclude such pairs from the calculation of forces using the pair list. This fact affects the subsequent SIMD vectorization.

\subsection{Sorting by Indices}
\begin{figure}[ht]
\includegraphics[width=8cm]{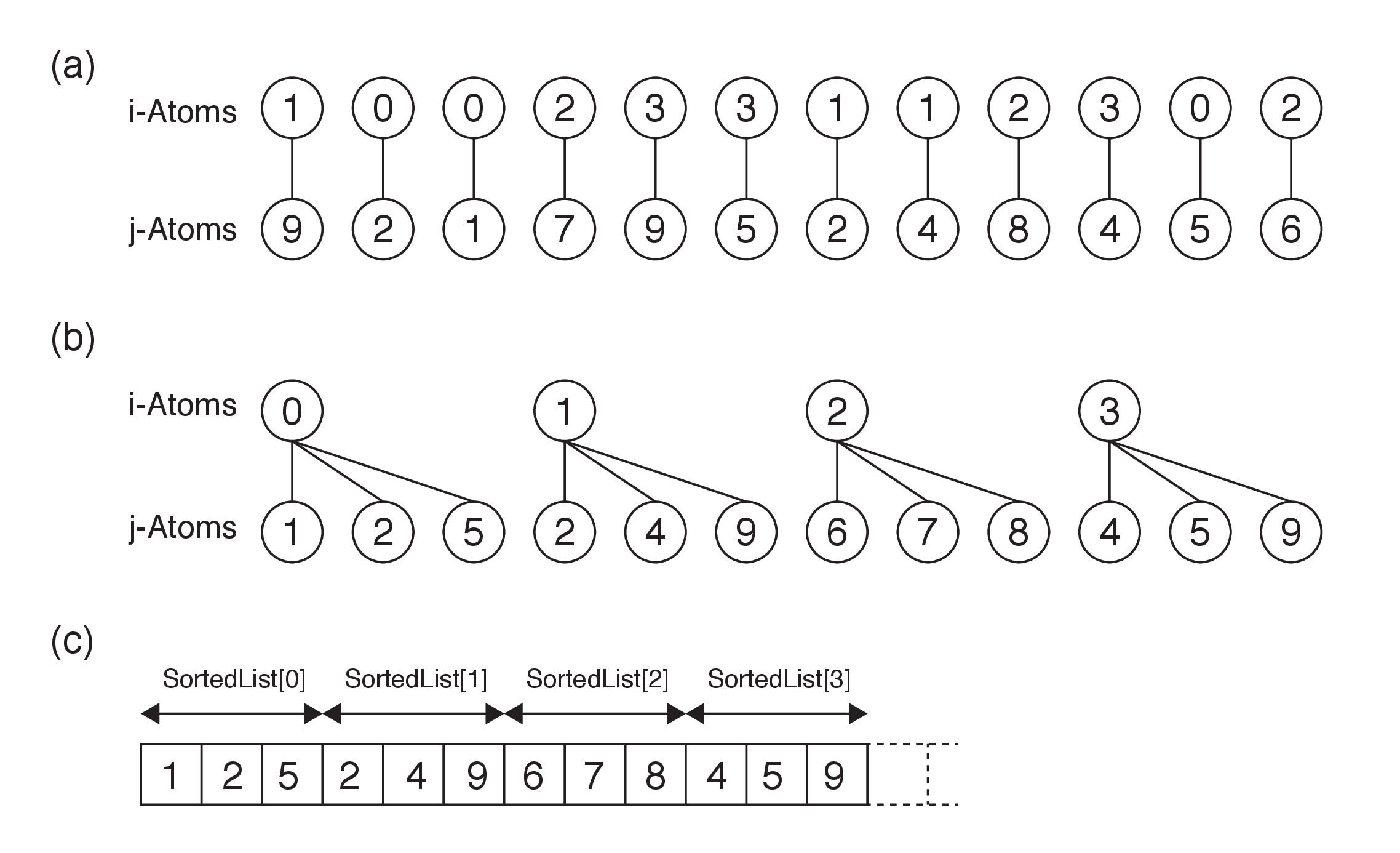}
\caption{
Data layout of a pair list.
(a) List of pairs.
(b) Sorted list. The $j$-atoms interacting with the same $i$-atom are grouped together.
We expect that the data of $i$-atoms are on registers.
(c) Array representation of (b). The array {\ttfamily SortedList[i]} denotes the list of $j$-atoms that interact with $i$-atom.
}
\label{fig:sorted_list}
\end{figure}

An interacting pair can be denoted by a pair of indices.
Therefore, the pair list is represented by a list of index pairs (see Fig.~\ref{fig:sorted_list}~(a)).
The pseudocode for calculating forces using the pair list is shown in Algorithm~\ref{alg:simple}.
The force-calculation kernel contains a single loop and 
it requires memory access both for $i$- and $j$-atoms.
Since the number of arithmetic operations is small as described before, 
the memory access becomes the bottleneck, resulting poor performance.
To reduce the number of memory accesses, we sort the pair list into group $j$-atoms
interacting with the same $i$-atom (see Fig.~\ref{fig:sorted_list}~(b)).
The array representation of the sorted list is shown in Fig.~\ref{fig:sorted_list}~(c).
The array \verb|SortedList[i]| denotes the array that stores the indices of $j$-atoms
interacting with $i$-atoms. This list can be constructed by a counting sort
and the computational complexity of the construction is $O(N$).
Using the sorted list, the force-calculation kernel contains double loops.
Then the information of the $i$-atom, which is denoted by the loop counter of the outer loop,
is stored in registers. Then only the memory access of $j$-atoms is required in the inner loop.
The pseudocode of the force calculation using the sorted list is shown in Algorithm~\ref{alg:calcforce_sortedlist}
Hereafter, we refer to this optimization to``Sorted".
This optimization works effectively on KNL as shown in Fig.~\ref{fig:scalar}.

\begin{algorithm}
\caption{Calculating the force in a simple manner}
\label{alg:simple}
\begin{algorithmic}[1]
\ForAll{ pairs $(i,j)$ in PairList}
\State Load $\vec{q}_i$
\State Load $\vec{q}_j$
\If{ $|\vec{q}_j -\vec{q}_i|  < r_\mathrm{c} $ }
\State Load $\vec{p}_i$
\State Load $\vec{p}_j$
\State Calculate force between $i$- and $j$-particles.
\State Update $\vec{p}_i$ and $\vec{p}_j$
\State Store $\vec{p}_i$
\State Store $\vec{p}_j$
\EndIf
\EndFor
\end{algorithmic}
\label{alg:pair_simple}
\end{algorithm} 

\begin{algorithm}
\caption{Calculating the force with a sorted list}
\label{alg:calcforce_sortedlist}
\begin{algorithmic}[1]
\For{ $i=0$ to $N-1$}
\State Load $\vec{q}_i$
\State Load $\vec{p}_i$
\ForAll{$j$  in SortedList[i]}
\State Load $\vec{q}_j$
\If{ $|\vec{q}_j -\vec{q}_i|  < r_\mathrm{c} $ }
\State Load $\vec{p}_j$
\State Calculate force between $i$- and $j$-particles.
\State Update $\vec{p}_i$ and $\vec{p}_j$
\State Store $\vec{p}_j$
\EndIf
\EndFor
\State Store $\vec{p}_i$
\EndFor
\end{algorithmic}
\label{alg:pair_sorted}
\end{algorithm} 

\subsection{Software Pipelining} \label{sec:swp}

\begin{figure}[ht]
\includegraphics[width=8cm]{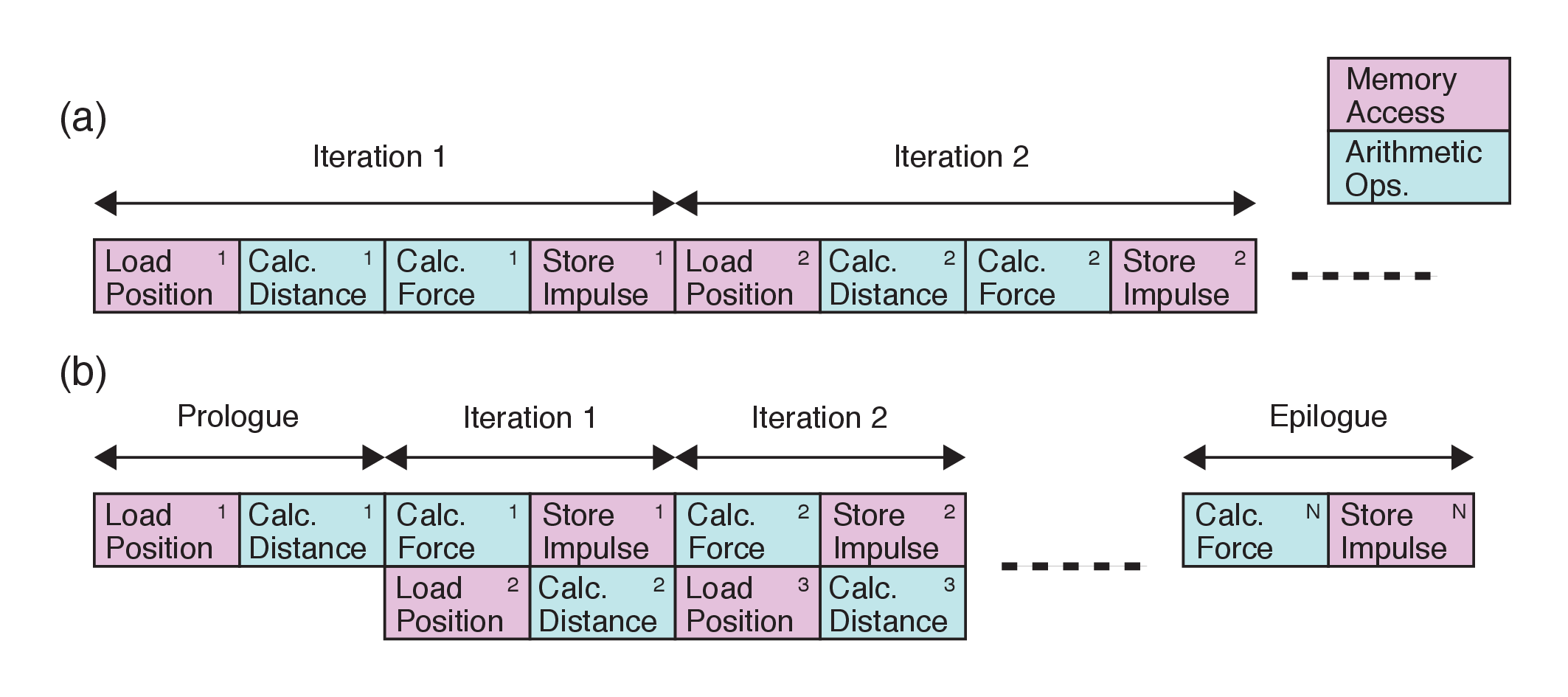}
\caption{
(Color online) Software pipelining. 
}
\label{fig:swp}
\end{figure}

Software pipelining (SWP) is one of the loop optimization techniques
that reforms the loop to reduce the execution time~\cite{Allan1995}.
While SWP is often used to hide the latency of instructions
so that pipelines are kept busy without stalling, we adopt this technique 
to increase the number of instructions per cycle (IPC) rather than to avoid pipeline hazards.
The structure of the force calculation has a double-loop form where 
the outer loop is for $i$-atoms and the inner loop is for $j$-atoms.
In the body of the inner loop, there are four different operations, $A, B, C$, and $D$, which are 
$A$: load the position of a $j$-atom,
$B$: calculate the distance between $i$- and $j$-atoms,
$C$: calculate the force and update the momenta
$D$: store the momentum of the $j$-atom.
Suppose $n$ is the number of $j$-atoms interacting with the $i$-atom,
then the inner loop has the form $\{ABCD\}^n$.
Since the four operations are mutually dependent, the operations 
should be performed sequentially (see Fig.~\ref{fig:swp}~(a)).
To increase parallelism, 
we change the inner loop from $\{ABCD\}^n$ to $AB\{CDAB\}^{n-1}CD$ (see Fig.~\ref{fig:swp}~(b))
so that the memory accesses and the arithmetic operations in the body of the inner loop are no longer mutually dependent.
Since the recent CPUs are superscalar, independent memory accesses and arithmetic operations can be executed simultaneously, increasing IPC.
Since the modification of the loop causes an increase in the number of instructions, the overall performance is improved when the increase in IPC is larger than the increase in the number of instructions.
The impact of this optimization technique, therefore, strongly depends on the environment, such as the CPU architecture, the version of the compiler, and so forth.

With the Intel 16.0.4 C++ compiler and on HSW, 
we find that SWP improves IPC by 55\% while the number of instructions increases by 22\%, 
and therefore, the total performance gain is about 21\%.
We confirm that SWP significantly improves the performance on SKL with
the Intel 17.0.6 C++ compiler. However, the gain in the performance vanishes when we use
the Intel 18.0.1 C++ compiler, since the performance without SWP is improved by using a later version of the compiler. The difference between the 17.0.6 and the 18.0.1 compilers is the vector optimization capability.
While the 17.0.6 compiler used xmm registers only, the 18.0.1 compilers vectorized the code
with using zmm registers. As the result, the code compiled by the 18.0.1 compilers is faster than that by the 17.0.6 compiler. Note that, we used the identical compiler options for both cases.
If we apply SWP by hand, then the loop was not vectorized since the loop body becomes too complicated.
However, the performance is improved since IPC increases.
Whether the performance improves by SWP depends on the balance between performance improvement by improving IPC and performance degradation due to inhibition of vectorization.
In SKL, the performances with and without SWP happened to be nearly the same.
However, the performance on KNL becomes worse with SWP.
This is because the auto-vectorization by the Intel compiler works efficiently for the code ``SoA+Sorted" on KNL.
The auto-vectorization by the compiler on KNL is discussed in Sec.~\ref{sec:autovec}.

\section{SIMD Vectorization with AVX2} \label{sec:avx2}

\begin{figure}[ht]
\includegraphics[width=8cm]{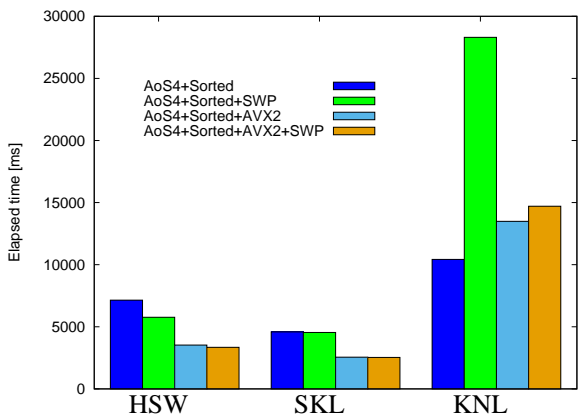}
\caption{
(Color online) Effects of optimizations with AVX2.
}
\label{fig:avx2}
\end{figure}

In this section, we describe the vectorization with AVX2 instructions.
We can use a 256-bit-width register with AVX2.
Therefore, four pairs of forces can ideally be calculated simultaneously.
Since the innermost loop of the MD code contains indirect accesses,
it is not straightforward to pack data from the memory to the register and to unpack data from the register to the memory.
Since the force calculations of the LJ potential are relatively light,
the packing and unpacking processes may be the bottleneck.
While AVX2 includes a gather instruction, it is very slow. Additionally, AVX2 does not have 
a scatter instruction.
Therefore, some ingenuities are required to pack and unpack the data.
The results for the codes vectorized with AVX2 are shown in Fig.~\ref{fig:avx2}.
In the following, we describe how to vectorize the force calculation with AVX2.

\subsection{AoS with Padding}

\begin{figure}[ht]
\includegraphics[width=8cm]{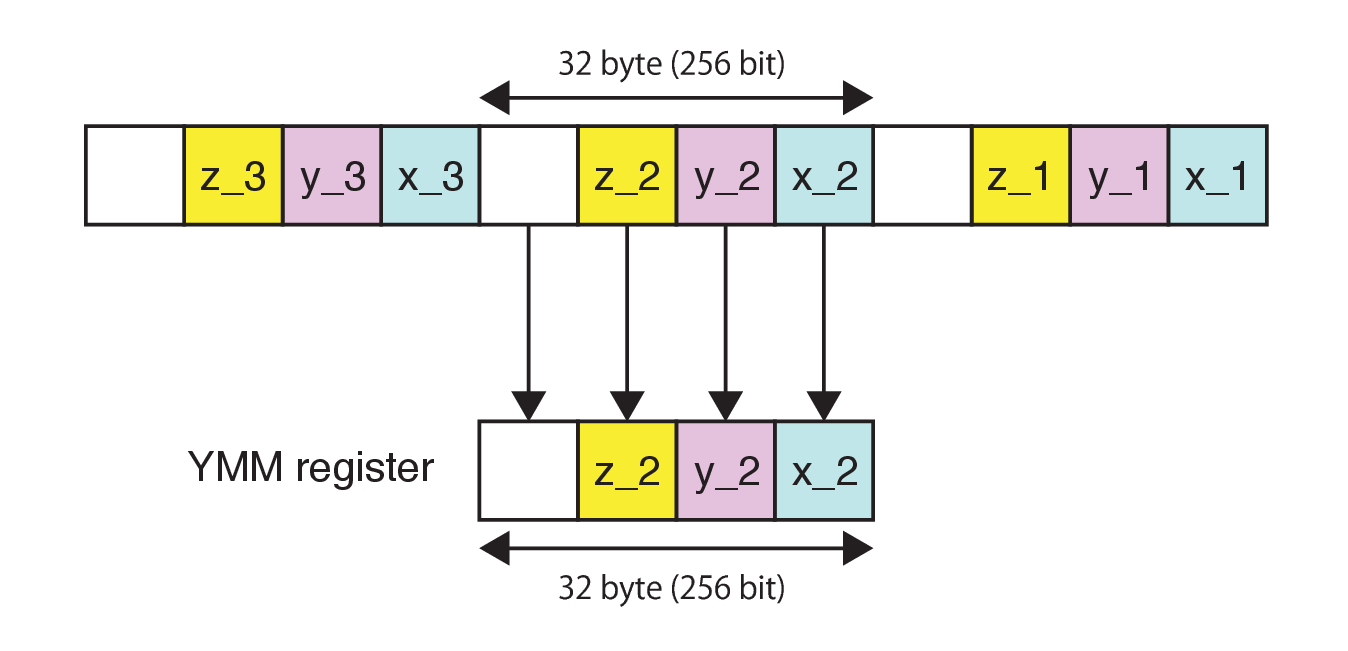}
\caption{
(Color online) Array of Structures (AoS) with padding.
This data layout allows us to load simultaneously three elements of coordinates to a YMM register.
}
\label{fig:aos4}
\end{figure}

\begin{figure}[ht]
\includegraphics[width=8cm]{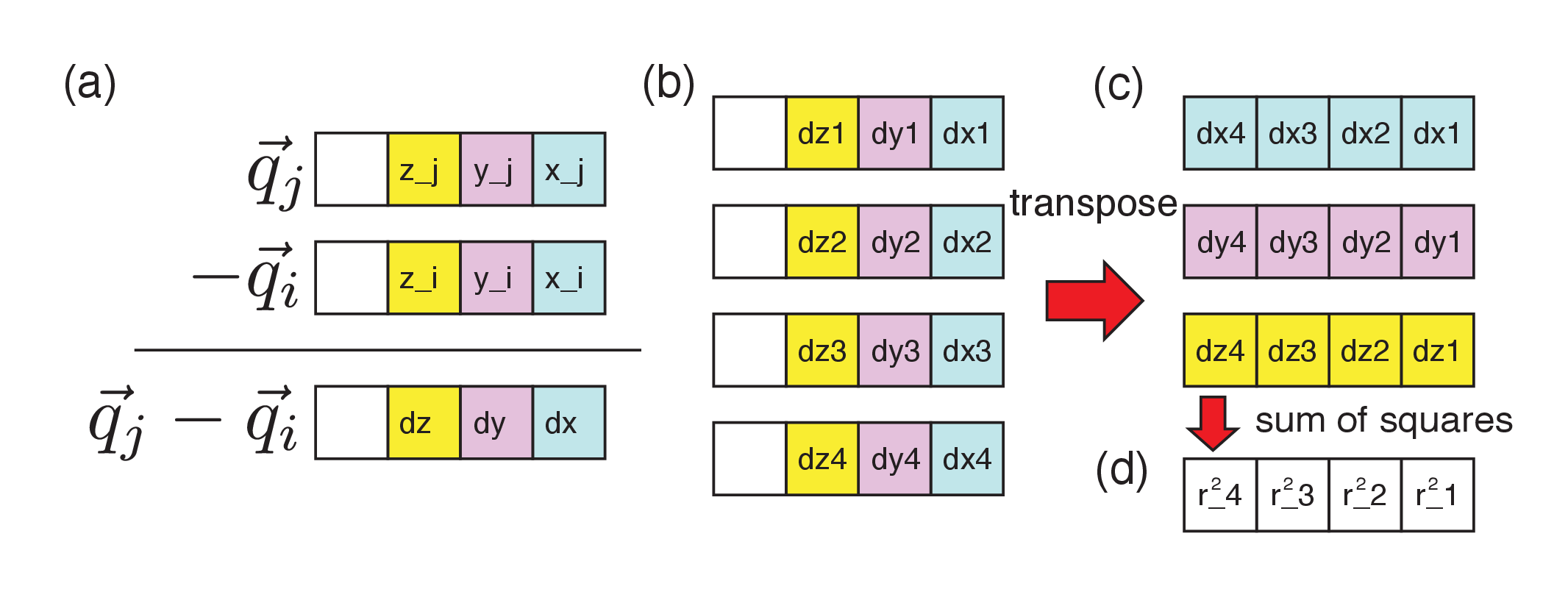}
\caption{
(Color online) Calculation of relative distances.
(a) By calculating the difference between the coordinates of the $i$- and $j$-atoms
stored in each YMM register, the relative position vector is stored in the YMM register.
(b) The calculation of the relative position vector is performed four times
for four $j$-atoms.
(c) The four YMM registers are transposed to obtain three registers.
(d) The sums of the squares of the three registers are calculated to obtain four squared relative distances of four pairs.
}
\label{fig:transpose}
\end{figure}

The key idea in the vectorization of the force calculation is to make the data structure the Array of Structures (AoS) with padding~\cite{Brown2015}.
The atom data are aligned so that the data structure has 256-bit boundaries.
Since the coordinates and momenta have three elements, 
one double precision floating number is inserted as a padding for every three elements (see Fig.~\ref{fig:aos4}).
We refer to this layout as ``AoS4" since each structure contains four elements including a padding.
This data structure has two advantages over SoA.
One is the cache efficiency. This data structure ensures that the data for a single atom
will be on a single cache line with 256-bit width.
The other is the memory-access efficiency.
Since the data are aligned on 256-bit boundaries,
the coordinates or momenta of a single atom can be moved to the YMM register by a single vector-load instruction (\verb|vmoveupd|). We find that the latter advantage is more effective for increasing the speed.
After loading the data of the coordinates into a register,
we calculate the relative coordinate vector between $i$- and $j$-atoms (Fig.~\ref{fig:transpose}~(a)).
Performing the above process four times, we obtain four registers containing relative coordinate vectors.
Then we transpose the four vectors to obtain three vectors (Fig.~\ref{fig:transpose}~(b)).
By calculating the sum of the squares of the vectors, we obtain a single register
that contains the relative distances of the four pairs (Fig.~\ref{fig:transpose}~(c)).

\begin{figure}[ht]
\includegraphics[width=8cm]{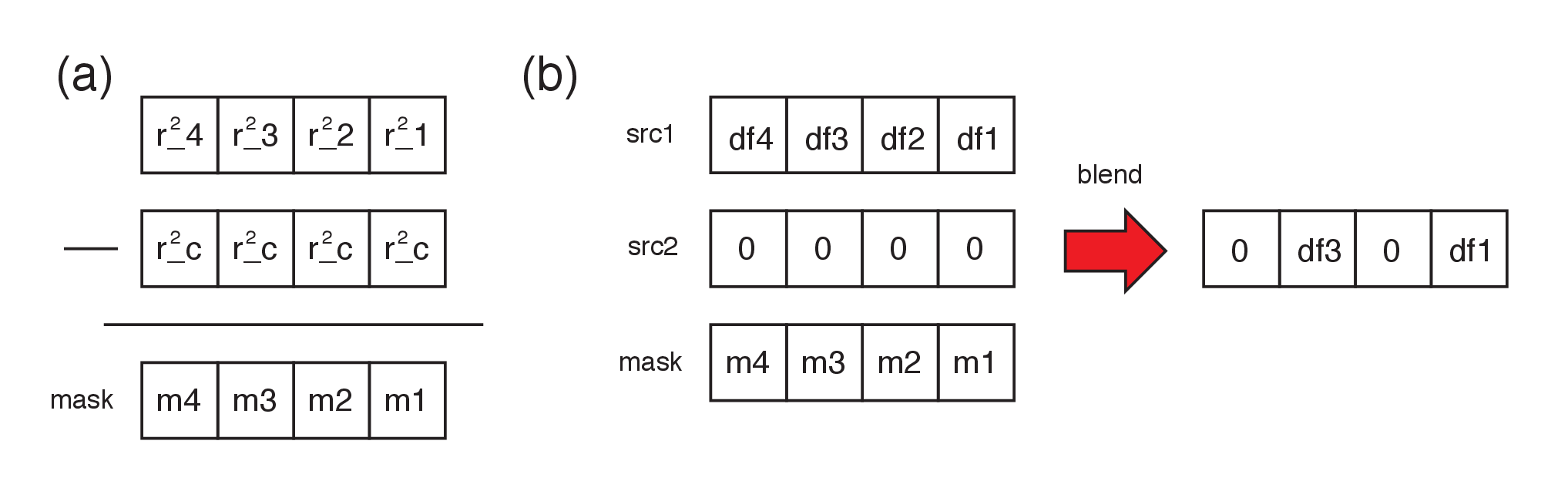}
\caption{
Conditional execution using masks.
The impulses between the pairs with distances greater than the truncation length
are set to zero.
}
\label{fig:mask}
\end{figure}

Since we adopt the bookkeeping method, the pair list contains the pairs with
distances greater than the truncation length.
This list can be treated by masking operations.
We first calculate impulses of all pairs.
Next, we set the impulses between pairs that are outside the interaction range to zero (see Fig.~\ref{fig:mask}).
Then we load the momenta of the $j$-atoms, update them using the calculated impulses, and write them back.

The results of the above vectorization are denoted by ``AoS4+Sorted+AVX2" in Fig.~\ref{fig:avx2}.
The increase in speed compared with the scalar code with SWP, \textit{i.e.}, the improvements from "AoS4+Sorted" to "AoS4+Sorted+AVX2", are 51\% on HSW and 44\% on SKL.
While the vectorization with AVX2 works effectively on HSW and SKL,
the vectorized code becomes significantly slower than the original code on KNL.
This is because automatic vectorization by the compiler works effectively with AVX-512 on KNL.
This issue will be discussed later.

\subsection{Software Pipelining}

We can apply the SWP technique to the vectorized code with AVX2 in a similar manner to the scalar code.
However, the performance improvements are moderate. 
While the performance is improved by 6\% on HSW, the effect on SKL is not significant.
Compared with the fastest scalar codes which is ``AoS4+Sorted+SWP",
the performance gains by vectorization are about 42\% on HSW and 44\% on KNL.
The performance on KNL deteriorates upon applying SWP.

\section{SIMD Vectorization with AVX-512} \label{sec:avx512}

\begin{figure}[ht]
\includegraphics[width=8cm]{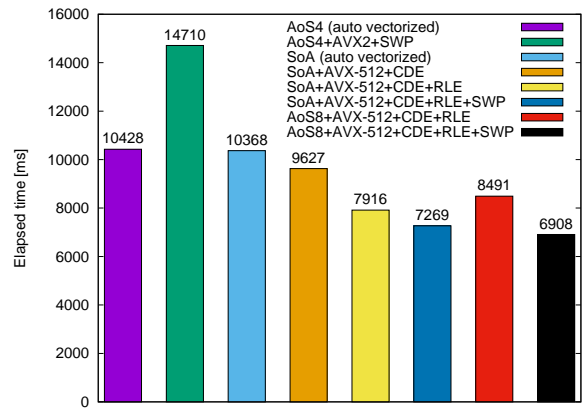}
\caption{
(Color online) Results on KNL. 
The meanings of the abbreviations in the figure are
AoS4 (Array of Structures with 4 members),
AoS8 (Array of Structures with 8 members),
SoA (Structure of Arrays),
CDE (collision detection elimination),
RLE (remainder loop elimination),
and SWP (software pipelining).
}
\label{fig:knl}
\end{figure}

In this section, we describe the vectorization with AVX-512 instructions.
The AVX-512 includes gather, scatter, and mask operations, which are useful
for the vectorization of the loop involving indirect access.
Here we describe the efficiency of the optimization on KNL.
The results are shown in Fig.~\ref{fig:knl}.
The results on SKL are discussed in the next section.

\subsection{Auto-Vectorization by Compiler} \label{sec:autovec}

While the Intel compiler cannot vectorize the benchmark program automatically with AVX2 without directives, it can vectorize the program with AVX-512.
Actually, the compiler vectorized the code in Algorithm~\ref{alg:calcforce_sortedlist} as follows.
\begin{enumerate}
\item Unroll the inner loop eight times.
\item Gather the positions and momenta of $j$-atoms with \verb|vgatherdpd|.
\item Calculate the forces and update the momenta of $j$-atoms.
\item Detect conflicts between the indices of $j$-atoms with \verb|vpconflictd|.
\item Scatter the updated momenta with \verb|vscatterdpd|.
\end{enumerate}
Here, the compiler inserted the codes for conflict detection since it did not know that there would be no conflict between the indices of $j$-atoms interacting with $i$-atoms. However, the penalty for inserting conflict detection is not expensive as shown later.
Inversely, the compiler cannot vectorize the code without conflict detection instructions.
In that sense, the conflict detection support in AVX-512 is crucial for auto-vectorization by the compiler as suggested by H\"ohnerbach \textit{et al}.~\cite{Hohnerbach2016}.

The code vectorized by the compiler with AVX-512 exhibits better performance than the code vectorized by hand with AVX2.
SWP reduces the performance since it interferes with the optimization by the compiler.

\subsection{AoS to SoA}

\begin{figure}[ht]
\includegraphics[width=8cm]{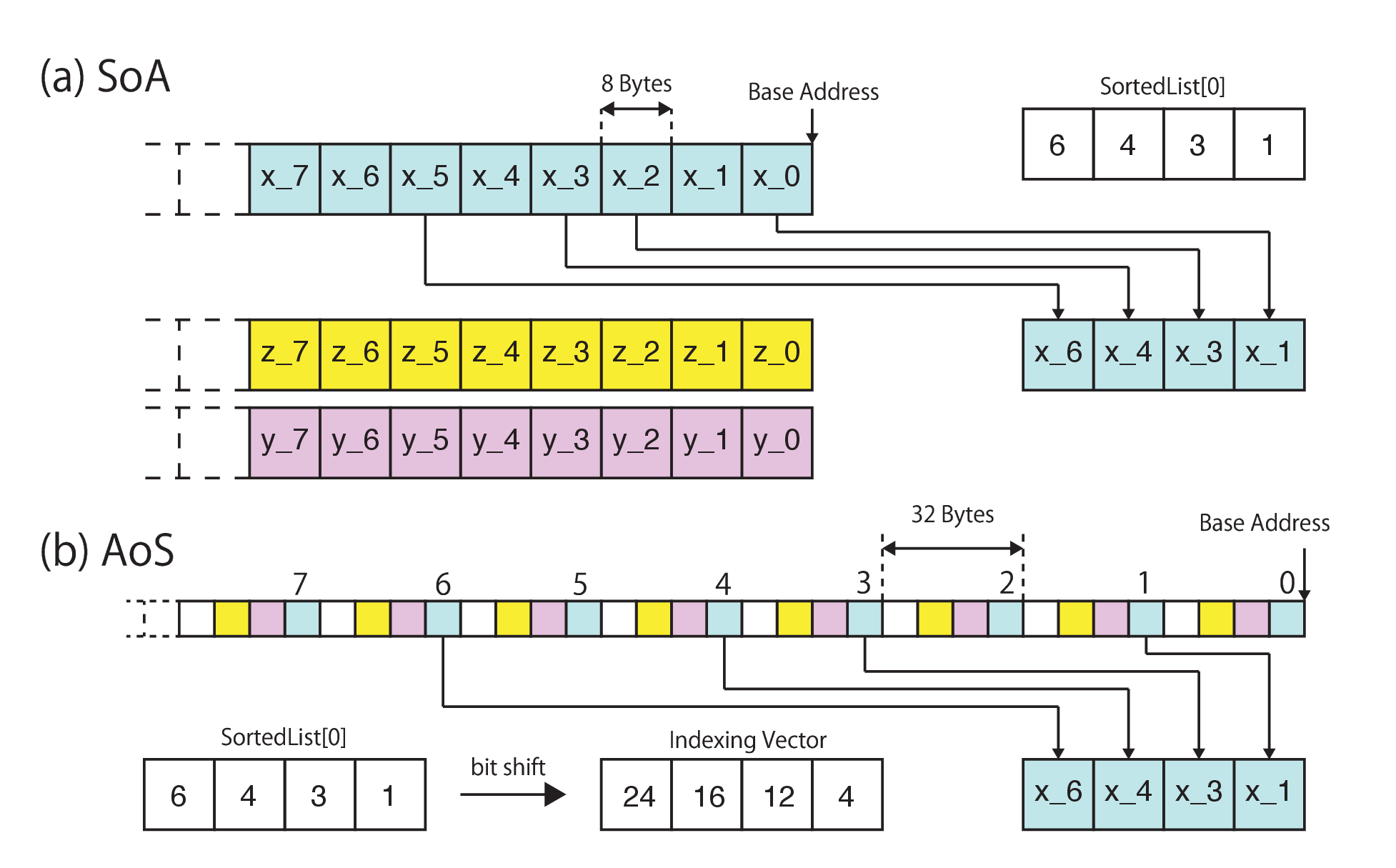}
\caption{
(Color online) Gather procedures. While eight double-precision floating elements can be gathered with AVX-512,
only four are shown for visibility. 
(a) Gather process for the SoA layout. By using 8 as the scale factor of the gather instruction, 
the data can be gathered by using the sorted list as is.
(b) Gather process for the AoS layout.
Since each element is separated by 32 bytes and the maximum scale factor is 8,
it is necessary to carry out the gather instruction using bit-shifted indices of the sorted list.
}
\label{fig:gather}
\end{figure}

Suppose $x_j$ is the $x$-coordinate of a $j$-atom.
The distance between $x_j$ and $x_{j+1}$ in memory is 8 bytes for SoA, while it is 32 bytes for AoS (see Fig.~\ref{fig:gather}). The gather (\verb|vgatherdpd|) and scatter (\verb|vscatterdpd|) instructions afford a scale factor that takes a values of 1, 2, 4, or 8.
When the data layout is SoA, then the indices of $j$-atoms can be directly used for gather/scatter instructions with a scale factor of 8. However, bit shifts are required for AoS, and therefore, the bit shifts impose some penalty on performance.
The efficiency of this optimization strongly depends on the code.
Actually, the improvement of the performance by changing the data layout is 1\% for the codes vectorized by the compiler.
However, it improves the performance by 7\% for the codes vectorized by hand with some additional optimizations, 
such as the remainder loop elimination and software pipelining, which are described later.

\subsection{Collision Detection Elimination}

As described above, the compiler inserted unnecessary codes for the conflict detection.
For the case of collision, the compiler prepares a code storing the momenta sequentially which will be never called. Since there are no conflicts between the indices, the conflict detection can be eliminated if we call gather and scatter instructions directly by using intrinsic functions.
By this conflict detection elimination (CDE), the performance is improved by 7\% (from "SoA+AVX-512+CDE" to "SoA+AVX-512+CDE+RLE").
Note that, it may be possible to tell the compiler that there are no conflicts using directives.
However, as far as we tried, the collision detection codes could not be eliminated by adding directives.

\subsection{Remainder Loop Elimination}

The number of iterations of the inner loop of the force calculation is equals the number of atoms existing within the interaction distance, which depends on the density of atoms.
For the benchmark condition adopted in the manuscript,
the number of $j$-atoms interacting with one $i$-atom is approximately 55 to 80.
If we unroll the inner loop eight times, the number of iterations
of the vectorized loop kernel becomes 6 to 10.
Since the number of iterations of the remaining loop is up to 7,
the computational time to process the remainder loop is comparable to the time required for the vectorized loop kernel.
Therefore, we eliminate the remainder loop by masking.
The pseudocode for the remainder loop elimination (RLE) is shown in Algorithm~\ref{alg:remainder_loop_elimination}.
The key idea is to prepare a vector of the loop counter ($\hat{d}$ in the pseudocode).
Then we can make a mask $\hat{m}_{\mathrm{loop}}$ for the remainder loop.
Note that we must also to make a mask $\hat{m}_{\mathrm{cutoff}}$ for the truncation of the interaction.
The forces are zeroed by a mask $\hat{m}_{\mathrm{total}}$ which is the logical conjunction
of $\hat{m}_{\mathrm{loop}}$ and $\hat{m}_{\mathrm{cutoff}}$.
The scatter of the momenta of $j$-atoms should be masked by $\hat{m}_{\mathrm{loop}}$
since there is a possibility of conflict due to the RLE.
RLE improves the performance by 18\% (from ``SoA+AVX-512+CDE" to ``SoA+AVX-512+CDE+RLE").

\begin{algorithm}
\caption{Remainder loop elimination}
\label{alg:loopopt}
\begin{algorithmic}[1]
\State $\hat{t} \leftarrow \{8,8,8,8,8,8,8,8\}$  
\State $\hat{v}_c \leftarrow \{r_c,r_c,r_c,r_c,r_c,r_c,r_c,r_c\}$  
\For{ $i=0$ to $N-1$}
\State $\hat{q}_i \leftarrow$ Position of $i$-Atom \Comment{Broadcast}
\State $\hat{d} \leftarrow \{7,6,5,4,3,2,1,0\}$  
\State $n \leftarrow$  $|\mathrm{SortedList[i]}|$
\State $k \leftarrow 0$
\State $\hat{n} \leftarrow \{n,n,n,n,n,n,n,n\}$
\While{$k < n$}
\State $\hat{q}_j \leftarrow$ Positions of $j$-Atoms \Comment{Gather}
\State $\hat{v}_d \leftarrow$ Distance between $\hat{q}_j$ and $\hat{q}_i$
\State Calculate forces between $i$- and $j$-atoms
\State $\hat{d} \leftarrow \hat{d} + \hat{t}$
\State $\hat{m}_{\mathrm{cutoff}} \leftarrow$ \verb|_mm512_cmp_pd_mask|$(\hat{v}_c,\hat{v}_r,\verb|_CMP_LE_OS|)$
\State $\hat{m}_{\mathrm{loop}} \leftarrow $\verb|_mm512_cmp_epi64_mask|$(\hat{d})$
\State $\hat{m}_{\mathrm{total}} \leftarrow $\verb|_mm512_kand|$(\hat{m}_{\mathrm{cutoff}}, \hat{m}_{\mathrm{loop}})$
\State Zero the forces with the mask $\hat{m}$
\State Store $\vec{p}_j$ with the mask $\hat{m}_{\mathrm{loop}}$
\State $k \leftarrow k + 8$
\EndWhile
\State Store $\vec{p}_j$
\EndFor
\end{algorithmic}
\label{alg:remainder_loop_elimination}
\end{algorithm} 

\subsection{Software Pipelining}

The performance of the vectorized kernel loop can be improved by 
the SWP technique described in Sec.~\ref{sec:swp}.
The performance improvement by SWP strongly depends on 
the optimization methods used together.
While SWP does not improve the efficiency of the code vectorized with AVX2,
it improves the performance of the code vectorized with AVX-512 by 9\% .

\subsection{AoS with Eight Elements}

\begin{figure}[ht]
\includegraphics[width=8cm]{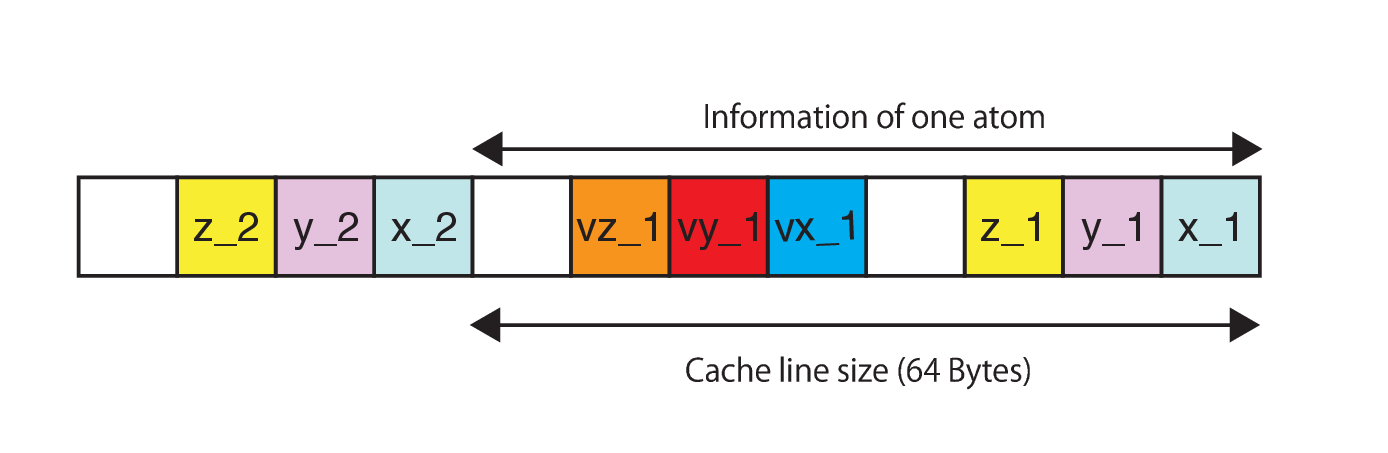}
\caption{
(Color online) AoS data layout with eight elements (AoS8).
Each structure has eight members, three of them are coordinates,
three of them are momenta, and two of them are padding, respectively.
The information of one atom is placed on one cache line.
}
\label{fig:aos8}
\end{figure}

From the viewpoint of the number of instructions, 
SoA is more advantageous than AoS since AoS requires additional instructions
to prepare the vector of indices for gather/scatter instructions.
However, the AoS data layout is more advantageous than that of SoA from the viewpoint of cache efficiency.
In the previous section, we introduced AoS with four elements: three of them were the coordinates or momenta of atoms
and the other one was padding. Here, we adopt AoS with eight elements so that all the information of one atom is placed within one cache line (see Fig.~\ref{fig:aos8}).
The optimal data layout depends on the optimization methods used together.
For the code applying CDE and RLE, SoA is superior to AoS8.
However, AoS8 exhibits better performance than SoA with SWP
since SWP significantly improves the performance of the code with AoS8.

\section{Results on Skylake} \label{sec:skl}

\begin{figure}[ht]
\includegraphics[width=8cm]{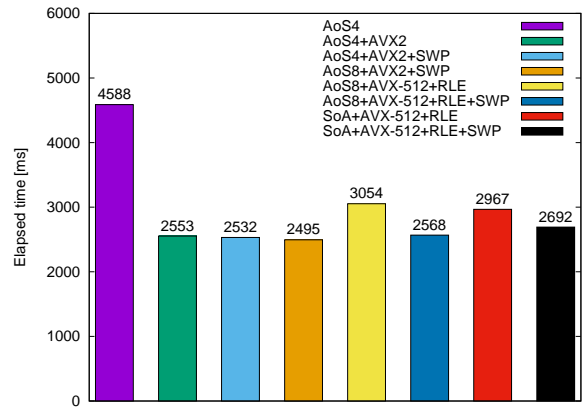}
\caption{
(Color online) Results on SKL.
The meanings of the abbreviations in the figure are
AoS4 (Array of Structures with 4 members),
AoS8 (Array of Structures with 8 members),
SoA (Structure of Arrays),
AVX2 (hand-vectorization with AVX2),
AVX-512 (hand-vectorization with AVX-512),
CDE (collision detection elimination),
RLE (remainder loop elimination),
and SWP (software pipelining).
}
\label{fig:skl}
\end{figure}

The results on SKL are shown in Fig.~\ref{fig:skl}.
The code with the AoS having four members and sorting by indices is denoted by ``AoS4".
This code is automatically vectorized by the Intel compiler using ZMM registers, but gather/scatter instructions
are not used and the performance is not satisfactory.
The results with ``AVX2" or ``AVX-512" are vectorized by hand using AVX2 or AVX-512, respectively.
We find that the code with AoS that is vectorized with AVX2 exhibits the best performance,
which is denoted by ``Aos8+AVX2+SWP" in the figure.
The performance improvement compared with the code automatically vectorized by the compiler,
\textit{i.e.}, from ``AoS4" to ``AoS8+AVX2+SWP", is about 46\%.

\section{Summary and Discussion} \label{sec:summary}

In this paper, we investigated the effect of combining different data layouts and
optimization methods used with vectorization for the truncated-LJ-force kernel
on different architectures, HSW, KNL, and SKL.
It was found that the choice of data layout is crucial for vectorization using AVX2 on HSW.
While AoS8 was found to exhibit the best performance on all the architectures studied in this work,
the differences in performance between AoS and SoA are less significant when the code is vectorized with AVX-512. 
We found that the data layout AoS8 may interfere with optimization by the compiler.
In the force-calculation kernel, the coordinates are only read,
but the momenta involve writing back. Since each structure in AoS8 includes both coordinates and momenta, the compiler
cannot determine which members can be modified.
Therefore, AoS4 is the optimal choice from the viewpoint of portability.
KNL and SKL support both AVX2 and AVX-512.
While the code vectorized with AVX-512 is much faster than that vectorized with AVX2 on KNL,
the code with AVX2 is slightly faster than that with AVX-512 on SKL.
Although SWP is an optimization technique that can be used together with vectorization with any instruction set, the gain in performance by SWP is not significant in this work.
However, we found that SWP works efficiently with another compilers such as GCC or with other architectures~\cite{Watanabe2011}.
Although its improvement in performance strongly depends on the optimization ability of the compiler,the SWP is worth considering as an optimization method to be used with vectorization.
In this work, we did not consider the compiler directives in detail since
it was easier to write codes with builtin-functions than searching for effective combinations of compiler directives. This situation may change for force calculations of more complex interactions.

While we performed vectorization faithfully to the original scalar code, the performance can be improved by using reciprocal approximations. Although reciprocal approximation instructions also exist in SSE and AVX2, more accurate instructions are available in AVX-512.
For example, \verb|vrcp28pd| computes an approximate reciprocal with 28-bit accuracy. Therefore, full precision can be obtained with the first-order correction.
However, \verb|vrcp28pd| is included in the AVX-512ER instruction set, which is supported by KNL but not SKL.
Instead, \verb|vrcp14pd| is available with SKL but it requires the second-order correction to obtain full precision.
Using these instructions, the performance of vectorization on SKL and KNL may be changed, which should be considered as a future issue.

While we were writing this paper, Intel announced discontinuance of Xeon Phi family~\cite{xeonphidiscontinued}.
Therefore, some part of this study became obsolete.
However, we think that the knowledge obtained in this study is still useful.
It is almost certain that the number of CPU cores and SIMD width will increase without increasing the operating frequency. Then the code vectorized with masked gather and scatter instructions are likely to be more efficient as it was on KNL, while the code vectorized with AVX2 was faster than that with AVX-512 on SKL.
In any case, it is difficult to predict the optimal combination of optimization techniques, and trial and error are necessary. We hope that our work will assist vectorizations and optimizations on the future architectures.

\section*{Acknowledgements}
The authors would like to thank S. Mitsunari and H. Noguchi for fruitful discussions.
This work was supported by JSPS KAKENHI Grant Number 15K05201 and
by the MEXT project as ``Exploratory Challenge on Post-K Computer" (Frontiers of Basic Science: Challenging the Limits).
The computations were carried out using the facilities of the Information Technology Center of the University of Tokyo and the Institute for Solid State Physics of the University of Tokyo.

\bibliography{ljsimd}

\begin{thebibliography}{21}
\expandafter\ifx\csname natexlab\endcsname\relax\def\natexlab#1{#1}\fi
\expandafter\ifx\csname bibnamefont\endcsname\relax
  \def\bibnamefont#1{#1}\fi
\expandafter\ifx\csname bibfnamefont\endcsname\relax
  \def\bibfnamefont#1{#1}\fi
\expandafter\ifx\csname citenamefont\endcsname\relax
  \def\citenamefont#1{#1}\fi
\expandafter\ifx\csname url\endcsname\relax
  \def\url#1{\texttt{#1}}\fi
\expandafter\ifx\csname urlprefix\endcsname\relax\def\urlprefix{URL }\fi
\providecommand{\bibinfo}[2]{#2}
\providecommand{\eprint}[2][]{\url{#2}}

\bibitem[{\citenamefont{Alder and Wainwright}(1957)}]{Alder1957}
\bibinfo{author}{\bibfnamefont{B.}~\bibnamefont{Alder}} \bibnamefont{and}
  \bibinfo{author}{\bibfnamefont{T.}~\bibnamefont{Wainwright}},
  \bibinfo{journal}{J. Chem. Phys.} \textbf{\bibinfo{volume}{27}},
  \bibinfo{pages}{1208} (\bibinfo{year}{1957}).

\bibitem[{\citenamefont{Germann and Kadau}(2008)}]{GERMANN2008}
\bibinfo{author}{\bibfnamefont{T.~C.} \bibnamefont{Germann}} \bibnamefont{and}
  \bibinfo{author}{\bibfnamefont{K.}~\bibnamefont{Kadau}},
  \bibinfo{journal}{Int. J. Mod. Phys. C} \textbf{\bibinfo{volume}{19}},
  \bibinfo{pages}{1315} (\bibinfo{year}{2008}).

\bibitem[{\citenamefont{Watanabe et~al.}(2013)\citenamefont{Watanabe, Suzuki,
  and Ito}}]{Watanabe2013}
\bibinfo{author}{\bibfnamefont{H.}~\bibnamefont{Watanabe}},
  \bibinfo{author}{\bibfnamefont{M.}~\bibnamefont{Suzuki}}, \bibnamefont{and}
  \bibinfo{author}{\bibfnamefont{N.}~\bibnamefont{Ito}},
  \bibinfo{journal}{Comput. Phys. Commun.} \textbf{\bibinfo{volume}{184}},
  \bibinfo{pages}{2775} (\bibinfo{year}{2013}).

\bibitem[{\citenamefont{Rupp}()}]{Rupp2018}
\bibinfo{author}{\bibfnamefont{K.}~\bibnamefont{Rupp}},
  \bibinfo{howpublished}{\url{https://github.com/karlrupp/microprocessor-trend-data}}.

\bibitem[{\citenamefont{Brown et~al.}(2015)\citenamefont{Brown, Carrillo,
  Gavhane, Thakkar, and Plimpton}}]{Brown2015}
\bibinfo{author}{\bibfnamefont{W.~M.} \bibnamefont{Brown}},
  \bibinfo{author}{\bibfnamefont{J.-M.~Y.} \bibnamefont{Carrillo}},
  \bibinfo{author}{\bibfnamefont{N.}~\bibnamefont{Gavhane}},
  \bibinfo{author}{\bibfnamefont{F.~M.} \bibnamefont{Thakkar}},
  \bibnamefont{and} \bibinfo{author}{\bibfnamefont{S.~J.}
  \bibnamefont{Plimpton}}, \bibinfo{journal}{Comput. Phys. Commun.}
  \textbf{\bibinfo{volume}{195}}, \bibinfo{pages}{95} (\bibinfo{year}{2015}).

\bibitem[{\citenamefont{Karpi{\'{n}}ski and McDonald}(2017)}]{Karpinski2017}
\bibinfo{author}{\bibfnamefont{P.}~\bibnamefont{Karpi{\'{n}}ski}}
  \bibnamefont{and} \bibinfo{author}{\bibfnamefont{J.}~\bibnamefont{McDonald}},
  in \emph{\bibinfo{booktitle}{Proceedings of the 8th International Workshop on
  Programming Models and Applications for Multicores and Manycores, PMAM’17}}
  (\bibinfo{publisher}{{ACM} Press}, \bibinfo{year}{2017}), pp.
  \bibinfo{pages}{21--28}.

\bibitem[{\citenamefont{H\"ohnerbach et~al.}(2016)\citenamefont{H\"ohnerbach,
  Ismail, and Bientinesi}}]{Hohnerbach2016}
\bibinfo{author}{\bibfnamefont{M.}~\bibnamefont{H\"ohnerbach}},
  \bibinfo{author}{\bibfnamefont{A.~E.} \bibnamefont{Ismail}},
  \bibnamefont{and}
  \bibinfo{author}{\bibfnamefont{P.}~\bibnamefont{Bientinesi}}, in
  \emph{\bibinfo{booktitle}{{SC}16: International Conference for High
  Performance Computing, Networking, Storage and Analysis}}
  (\bibinfo{publisher}{{IEEE}}, \bibinfo{year}{2016}).

\bibitem[{\citenamefont{Abraham et~al.}(2015)\citenamefont{Abraham, Murtola,
  Schulz, P{\'{a}}ll, Smith, Hess, and Lindahl}}]{Abraham2015}
\bibinfo{author}{\bibfnamefont{M.~J.} \bibnamefont{Abraham}},
  \bibinfo{author}{\bibfnamefont{T.}~\bibnamefont{Murtola}},
  \bibinfo{author}{\bibfnamefont{R.}~\bibnamefont{Schulz}},
  \bibinfo{author}{\bibfnamefont{S.}~\bibnamefont{P{\'{a}}ll}},
  \bibinfo{author}{\bibfnamefont{J.~C.} \bibnamefont{Smith}},
  \bibinfo{author}{\bibfnamefont{B.}~\bibnamefont{Hess}}, \bibnamefont{and}
  \bibinfo{author}{\bibfnamefont{E.}~\bibnamefont{Lindahl}},
  \bibinfo{journal}{{SoftwareX}} \textbf{\bibinfo{volume}{1-2}},
  \bibinfo{pages}{19} (\bibinfo{year}{2015}).

\bibitem[{\citenamefont{Edwards et~al.}(2014)\citenamefont{Edwards, Trott, and
  Sunderland}}]{Edwards2014}
\bibinfo{author}{\bibfnamefont{H.~C.} \bibnamefont{Edwards}},
  \bibinfo{author}{\bibfnamefont{C.~R.} \bibnamefont{Trott}}, \bibnamefont{and}
  \bibinfo{author}{\bibfnamefont{D.}~\bibnamefont{Sunderland}},
  \bibinfo{journal}{J. Parallel Distrib. Comput.}
  \textbf{\bibinfo{volume}{74}}, \bibinfo{pages}{3202} (\bibinfo{year}{2014}).

\bibitem[{\citenamefont{Phillips et~al.}(2005)\citenamefont{Phillips, Braun,
  Wang, Gumbart, Tajkhorshid, Villa, Chipot, Skeel, Kal{\'{e}}, and
  Schulten}}]{Phillips2005}
\bibinfo{author}{\bibfnamefont{J.~C.} \bibnamefont{Phillips}},
  \bibinfo{author}{\bibfnamefont{R.}~\bibnamefont{Braun}},
  \bibinfo{author}{\bibfnamefont{W.}~\bibnamefont{Wang}},
  \bibinfo{author}{\bibfnamefont{J.}~\bibnamefont{Gumbart}},
  \bibinfo{author}{\bibfnamefont{E.}~\bibnamefont{Tajkhorshid}},
  \bibinfo{author}{\bibfnamefont{E.}~\bibnamefont{Villa}},
  \bibinfo{author}{\bibfnamefont{C.}~\bibnamefont{Chipot}},
  \bibinfo{author}{\bibfnamefont{R.~D.} \bibnamefont{Skeel}},
  \bibinfo{author}{\bibfnamefont{L.}~\bibnamefont{Kal{\'{e}}}},
  \bibnamefont{and} \bibinfo{author}{\bibfnamefont{K.}~\bibnamefont{Schulten}},
  \bibinfo{journal}{J. Comput. Chem.} \textbf{\bibinfo{volume}{26}},
  \bibinfo{pages}{1781} (\bibinfo{year}{2005}).

\bibitem[{lj_()}]{lj_simd}
\bibinfo{howpublished}{\url{https://github.com/kaityo256/lj_simd}}.

\bibitem[{\citenamefont{Watanabe et~al.}(2011)\citenamefont{Watanabe, Suzuki,
  and Ito}}]{Watanabe2011}
\bibinfo{author}{\bibfnamefont{H.}~\bibnamefont{Watanabe}},
  \bibinfo{author}{\bibfnamefont{M.}~\bibnamefont{Suzuki}}, \bibnamefont{and}
  \bibinfo{author}{\bibfnamefont{N.}~\bibnamefont{Ito}},
  \bibinfo{journal}{Prog. Theor. Phys.} \textbf{\bibinfo{volume}{126}},
  \bibinfo{pages}{203} (\bibinfo{year}{2011}).

\bibitem[{\citenamefont{Stoddard and Ford}(1973)}]{Stoddard1973}
\bibinfo{author}{\bibfnamefont{S.~D.} \bibnamefont{Stoddard}} \bibnamefont{and}
  \bibinfo{author}{\bibfnamefont{J.}~\bibnamefont{Ford}},
  \bibinfo{journal}{Phys. Rev. A} \textbf{\bibinfo{volume}{8}},
  \bibinfo{pages}{1504} (\bibinfo{year}{1973}).

\bibitem[{\citenamefont{Broughton and Gilmer}(1983)}]{Broughton1983}
\bibinfo{author}{\bibfnamefont{J.~Q.} \bibnamefont{Broughton}}
  \bibnamefont{and} \bibinfo{author}{\bibfnamefont{G.~H.}
  \bibnamefont{Gilmer}}, \bibinfo{journal}{J. Chem. Phys.}
  \textbf{\bibinfo{volume}{79}}, \bibinfo{pages}{5095} (\bibinfo{year}{1983}).

\bibitem[{\citenamefont{Holian et~al.}(1991)\citenamefont{Holian, Voter,
  Wagner, Ravelo, Chen, Hoover, Hoover, Hammerberg, and Dontje}}]{Holian1991}
\bibinfo{author}{\bibfnamefont{B.~L.} \bibnamefont{Holian}},
  \bibinfo{author}{\bibfnamefont{A.~F.} \bibnamefont{Voter}},
  \bibinfo{author}{\bibfnamefont{N.~J.} \bibnamefont{Wagner}},
  \bibinfo{author}{\bibfnamefont{R.~J.} \bibnamefont{Ravelo}},
  \bibinfo{author}{\bibfnamefont{S.~P.} \bibnamefont{Chen}},
  \bibinfo{author}{\bibfnamefont{W.~G.} \bibnamefont{Hoover}},
  \bibinfo{author}{\bibfnamefont{C.~G.} \bibnamefont{Hoover}},
  \bibinfo{author}{\bibfnamefont{J.~E.} \bibnamefont{Hammerberg}},
  \bibnamefont{and} \bibinfo{author}{\bibfnamefont{T.~D.}
  \bibnamefont{Dontje}}, \bibinfo{journal}{Phys. Rev. A}
  \textbf{\bibinfo{volume}{43}}, \bibinfo{pages}{2655} (\bibinfo{year}{1991}).

\bibitem[{\citenamefont{Quentrec and Brot}(1973)}]{Quentrec1973}
\bibinfo{author}{\bibfnamefont{B.}~\bibnamefont{Quentrec}} \bibnamefont{and}
  \bibinfo{author}{\bibfnamefont{C.}~\bibnamefont{Brot}}, \bibinfo{journal}{J.
  Comput. Phys.} \textbf{\bibinfo{volume}{13}}, \bibinfo{pages}{430}
  (\bibinfo{year}{1973}).

\bibitem[{\citenamefont{Hockney and Eastwood}(1988)}]{Hockney1988}
\bibinfo{author}{\bibfnamefont{R.~W.} \bibnamefont{Hockney}} \bibnamefont{and}
  \bibinfo{author}{\bibfnamefont{J.~W.} \bibnamefont{Eastwood}},
  \emph{\bibinfo{title}{Computer Simulation Using Particles}}
  (\bibinfo{publisher}{Taylor \& Francis Ltd}, \bibinfo{year}{1988}), ISBN
  \bibinfo{isbn}{0852743920}.

\bibitem[{\citenamefont{Verlet}(1967)}]{Verlet1967}
\bibinfo{author}{\bibfnamefont{L.}~\bibnamefont{Verlet}},
  \bibinfo{journal}{Phys. Rev.} \textbf{\bibinfo{volume}{159}},
  \bibinfo{pages}{98} (\bibinfo{year}{1967}).

\bibitem[{\citenamefont{Isobe}(1999)}]{Isobe1999}
\bibinfo{author}{\bibfnamefont{M.}~\bibnamefont{Isobe}}, \bibinfo{journal}{Int.
  J. Mod. Phys. C} \textbf{\bibinfo{volume}{10}}, \bibinfo{pages}{1281}
  (\bibinfo{year}{1999}).

\bibitem[{\citenamefont{Allan et~al.}(1995)\citenamefont{Allan, Jones, Lee, and
  Allan}}]{Allan1995}
\bibinfo{author}{\bibfnamefont{V.~H.} \bibnamefont{Allan}},
  \bibinfo{author}{\bibfnamefont{R.~B.} \bibnamefont{Jones}},
  \bibinfo{author}{\bibfnamefont{R.~M.} \bibnamefont{Lee}}, \bibnamefont{and}
  \bibinfo{author}{\bibfnamefont{S.~J.} \bibnamefont{Allan}},
  \bibinfo{journal}{ACM Comput. Surv.} \textbf{\bibinfo{volume}{27}},
  \bibinfo{pages}{367} (\bibinfo{year}{1995}).

\bibitem[{xeo()}]{xeonphidiscontinued}
\bibinfo{howpublished}{\url{http://qdms.intel.com/dm/i.aspx/9C54A9A7-BF37-4496-B268-BD2746EA54D3/PCN116378-00.pdf}}.

\end{thebibliography}

\end{document}